\documentclass[aps,showpacs,preprintnumbers,amsmath,amssymb]{revtex4}

\UseRawInputEncoding

\usepackage{graphicx}
\usepackage{subfig}
\usepackage{esint}

\usepackage{amsmath,amssymb,amsfonts}

\begin{document}
\baselineskip=0.8 cm
\captionsetup{justification=raggedright,singlelinecheck=false}
\title{Polarized image of a rotating black hole surrounded by a cold dark matter halo}

\author{Xin Qin$^{1}$,
Songbai Chen$^{1,2}$\footnote{Corresponding author: csb3752@hunnu.edu.cn}, Zelin Zhang$^{1}$,
Jiliang Jing$^{1,2}$ \footnote{jljing@hunnu.edu.cn}}
\affiliation{ $ ^1$ Department of Physics,  Institute of Interdisciplinary Studies, Synergetic Innovation Center for Quantum Effects and Applications, Hunan
Normal University,  Changsha, Hunan 410081, People's Republic of China
\\
$ ^2$Center for Gravitation and Cosmology, College of Physical Science and Technology, Yangzhou University, Yangzhou 225009, People's Republic of China}

\begin{abstract}
\baselineskip=0.6 cm
\begin{center}
{\bf Abstract}
\end{center}

We have studied the polarized image of an equatorial emitting ring around a rotating black hole surrounded by a cold dark matter (CDM) halo. Results show that the CDM halo density has the similar effects of the halo's characteristic radius on the polarized image for the black hole. The effects of the CDM halo on the polarized image depend on the magnetic field configuration, the fluid velocity and the observed inclination.
With the increase of the CDM halo parameters, the observed polarization intensity decreases  when the magnetic field lies in equatorial plane, but in the case where the magnetic field is perpendicular to the equatorial plane, the change of the observed polarization intensity with CDM halo also depends on the position of the emitting point in the ring. The change of the electric vector position angle (EVPA) with the CDM halo becomes more complicated. Our results also show that the influence
of the CDM halo on the polarized image is generally small, which are consistent with the effects of dark matter halo on  black hole shadows.  These results could help to further understand dark matter from black hole images.

\end{abstract}

 \maketitle
\newpage
\section{Introduction}

The releasing of black hole images of M87* \cite{EHT1,EHT2,EHT3,EHT4,EHT5,EHT6} and Sgr A*\cite{EHT7} by the Event Horizon Telescope (EHT) collaboration,  together with the corresponding polarized patterns \cite{EHT8,EHT9},  means that the observational black hole astronomy has been entered an new era of rapid progress. It is helpful to confirm the existence of black holes, to test general relativity and to explore matter distribution around black holes. The polarized image of M87* \cite{EHT8,EHT9} revealed that there is a strong magnetic field around the black hole. It is beneficial to understand the process of accretion and the generation of jet in the vicinity of black holes in universe. Therefore,  a lot of polarized images of black holes have been studied in both theoretical and experimental aspects \cite{PZ1,PZ2,PZ3,PZ4,PZ5,PZ6,PZ7,PZ8,PZ9,PZJG1,PZJG2,PZJG3,PZJG4,PZJG5,PZJG6,PZJG7,PZJG8,PZJG9,PZJG10,PZJG11,
PZJG12,PZJG121,Himwich,extra1,extra2}.

In general, numerical simulations are applied to obtain an accurate description of polarization image of a black hole. However,  they are computationally expensive because of the broad parameter surveys and the complicated couplings among astrophysical and relativistic effects. Recently, a simple model of an equatorial ring of magnetized fluid has been proposed to study the polarized images of synchrotron emission around Schwarzschild \cite{PZ1} and Kerr black hole \cite{PZ2}. Although only the emission from a single radius is considered in this model, it can clearly reveal the dependence of the polarization signatures on the magnetic field configuration, the black hole spin and the observer inclination. In addition,  the studies \cite{PZ1,PZ2} also indicate that the ring model image is broadly consistent with the polarization
morphology of the EHT image,  even though there is a high fractional polarization after blurring.
Thus, the simple ring model has been recently applied
to study the polarized image of an equatorial emitting ring in various spacetimes, such as, a 4D Gauss-Bonnet black hole spacetime \cite{PZ3}, a rotating black hole spacetime in the STVG-MOG theory \cite{PZ4}, regular black hole spacetimes \cite{PZ7}, a Schwarzschilld-Melvin black hole spacetime \cite{PZ8}, and so on.

Dark matter is a theoretical model to explain the discrepancy between the observed dynamics and the amount of luminous matter. However, the nature of dark matter is still open.
Dark matter is assumed to be an invisible matter and has feeble couplings with the common visible matter at most. Despite extensive observational data  supporting its presence on a large scale \cite{DM1,DM2,DM3}, dark matter has not been  directly detected  at present. The CDM model is one of the most popular  dark matter models \cite{DM4,DM5}, which has a good consistency with the observation of large scale structure of cosmology. The spacetime metric of Schwarzschild black hole and Kerr black hole surrounded by a CDM halo were studied in \cite{CDM1} and the effects of the CDM halo on the quasinormal modes
have also been studied in the spherical symmetric black hole spacetime \cite{CDM4}. The studies of black hole shadows \cite{CDM2,CDM201} show that the effects of the CDM halo on shadows are generally small and only become significant when the CDM  parameter $k=\rho_c \tilde{R}^3$ increases to order of magnitude of $10^7$.
However, they are very important for understanding dark matter from black hole shadows.
Along this line, we here will study the polarized image of the equatorial emitting ring around a rotating black hole surrounded by a CDM halo and probe the effects of the CDM halo on the polarization image.

The paper is organized as follows: Section II briefly introduces the rotating black hole surrounded by a CDM halo and presents the calculation formulas for the observed polarization vector in the image of an emitting ring in this spacetime. Section III presents the polarization images of the synchrotron emitting ring and probes the effects of the CDM halo. Finally, this paper ends with a summary.

\section{Observed polarization field in a rotating black hole spacetime surrounded by dark matter halo}

The metric analytical form of a rotating black hole surrounded by a CDM halo was obtained in \cite{CDM1}. In the Boyer-Lindquist coordinates, it has the form
\begin{equation}\label{metric1}
\begin{split}
ds^2=-&\left[1-\frac{r^2-f(r)r^2}{\rho^2}\right]dt^2+\frac{\rho^2}{\Delta}dr^2+\frac{2\left[r^2-f(r)r^2\right]a\sin^2\theta}{\rho^2}d\phi{dt}+\\
&\quad\quad\quad\rho^2d\theta^2+\frac{\sin\theta^2}{\rho^2}\left[\left(r^2+a^2\right)^2-a^2\Delta\sin\theta^2\right]d\phi^2,
\end{split}
\end{equation}
with
\begin{eqnarray}\label{Metric2}
&&f(r)=\left(1+\frac{r}{\tilde{R}}\right)^{-\frac{8\pi{G}\rho_c\tilde{R}^3}{c^2r}}-\frac{2GM}{rc^2},
\quad\quad\quad \Delta=r^2f(r)+a^2,\quad\quad\quad \rho^2=r^2+a^2\cos\theta^2,
\end{eqnarray}
where $M$ is the mass of the black hole, $G$ and $c$ are Newtonian gravitational constant and the speed of light,  respectively. $\rho_c$ is the density of the halo at the moment when the dark matter halo collapsed and $\tilde{R}$ is the corresponding characteristic radius. According to the observations on the Sgr A$^{\ast}$ \cite{CDM2}, the best fit values of the rotation curve of the Galaxy are $\rho_c$ = 1.936$\times10^7$ M$_{\odot}$kpc$^{-3}$ and $\tilde{R} = 17.46$ kpc. The metric can reduce to the Kerr metric in the limit $\rho_c=0$. Here, we set $G = 1$ and $c = 1$.

In the spacetime of a rotating black hole surrounded by a CDM halo (\ref{metric1}),  the geodesic equation for photons can be expressed as
\begin{eqnarray}\label{geodesic}
&&\frac{\rho^2}{E}p^t=\frac{r^2+a^2}{\Delta}\left(r^2+a^2-a\lambda\right)+a\left(\lambda-a\sin\theta^2\right), \\ \nonumber
&&\frac{\rho^2}{E}p^\phi=\frac{a}{\Delta}\left(r^2+a^2-a\lambda\right)+\frac{\lambda}{\sin\theta^2}-a, \\ \nonumber
&&\frac{\rho^2}{E}p^r=\pm_r\sqrt{\mathcal{R}(r)}, \\ \nonumber
&&\frac{\rho^2}{E}p^\theta=\pm_\theta\sqrt{\Theta(\theta)}.
\end{eqnarray}
The conserved quantities $\lambda$ and $\eta$ are the energy-rescaled angular momentum parallel to the axis of symmetry and Carter constant, respectively. The radial potential $\mathcal{R}(r)$ and the angular potential $\Theta(\theta)$ take the form
\begin{eqnarray}\label{potential}
&&\mathcal{R}(r)=\left(r^2+a^2-a\lambda\right)^2-\Delta\left[\eta+(a-\lambda)^2\right],  \\ \nonumber
&&\Theta(\theta)=\eta+a^2\cos\theta^2-\lambda^2\cot\theta^2.
\end{eqnarray}
Utilizing the null geodesic Eqs.\eqref{geodesic}, the coordinates $(x,y)$ for the photon's arrival position on the observer's screen  can be obtained as
\begin{eqnarray}\label{xy}
x=-\frac{\lambda}{\sin\theta_o}, \quad\quad\quad  y=\pm_o\sqrt{\Theta(\theta)},
\end{eqnarray}
where  $\theta_o$ is the observed inclination angle from the normal direction of the accretion disk. The radial and angle integrals of the photon trajectories from the initial position $(r_s,\theta_s)$ to the final initial $(r_o,\theta_o)$ can be expressed as \cite{PZ2,Math1}
\begin{eqnarray}\label{integra}
I_r\equiv\fint_{r_s}^{r_o}\frac{dr}{\pm_r\sqrt{\mathcal{R}(r)}}=\fint_{\theta_s}^{\theta_o}\frac{d\theta}{\pm_r\sqrt{\Theta(\theta)}}\equiv{G_\theta}. \end{eqnarray}
The slash in the path integral denotes that the sign of $\pm_r$ or $\pm_\theta$ changes when the photon passes through at the radial or angular turning point. For a photon's trajectory with $m$ turning points, the angular path integral can be written as
\begin{eqnarray}\label{xy1}
G_\theta^m=\frac{1}{\sqrt{-u_{-}a^2}}\left[2mK\left(\frac{u_+}{u_-}\right)-sign(\beta)F_o\right],
\end{eqnarray}
where
\begin{eqnarray}\label{xy03}
F_o=F\left(\arcsin\frac{\cos\theta_o}{\sqrt{u_+}}\Big|\frac{u_+}{u_-}\right),  \quad\quad u_\pm=\Delta_\theta\pm\sqrt{\Delta_\theta^2+\frac{\eta}{a^2}},  \quad\quad  \Delta_\theta=\frac{1}{2}\left(1-\frac{\eta+\lambda^2}{a^2}\right).
\end{eqnarray}
Combining Eqs.\eqref{xy}, Eqs.\eqref{integra} and Eqs.\eqref{xy1}, one can numerically get the set of celestial coordinates $(x,y)$ for a photon emitted on an equatorial ring with a radius $r_s$. The four-momentum of the photon at the source $(r_s, \theta_s=\frac{\pi}{2})$ is given by
\begin{eqnarray}\label{momentum1}
p_t=-1,  \quad\quad\quad\quad  p_r=\pm_r\frac{\sqrt{\mathcal{R}(r_s)}}{\Delta_s},  \quad\quad\quad\quad  p_\theta=\pm_s\sqrt{\eta},   \quad\quad\quad\quad  p_\phi=\lambda.
\end{eqnarray}
Here, the sign of $p_\theta$ at the source $\pm_s = (-1)^m\pm_o$ and the sign of $\pm_r$ can be computed by a semi-analytic method \cite{PZ2,Math1}. The four-momentum $p^\mu$ of photon at the source can be obtained as
\begin{eqnarray}\label{momentum03}
&&p^t=\frac{1}{r_s^2}\left\{\frac{r_s^2+a^2}{\Delta_s}\left(r_s^2+a^2-a\lambda\right)+a(\lambda-a)\right\}, \quad\quad  p^r=\pm_r\frac{1}{r_s^2}\sqrt{\mathcal{R}(r_s)},  \\ \nonumber
&&p^\phi=\frac{1}{r_s^2}\left\{\frac{a}{\Delta_s}\left(r_s^2+a^2-a\lambda\right)+\lambda-a\right\},   \quad\quad\quad\quad\quad p^\theta=\pm_s\frac{\sqrt{\eta}}{r_s^2}.
\end{eqnarray}
The emitted ring around a rotating black hole surrounded by a CDM halo (\ref{metric1}) is assumed to lie in the equatorial plane $(\theta_s=\frac{\pi}{2})$. In the local orthonormal zero-angular-momentum-observer (ZAMO) frame of the point $P$, the boosting velocity of the boosted emitter is assumed to be in the $r$-$\phi$ plane and has a form
\begin{eqnarray}\label{boost1}
\vec{\beta}=\beta_\nu\left(\cos\chi\left(\hat{r}\right)+\sin\chi(\hat{\phi})\right).
\end{eqnarray}
In the boosted orthonormal frame, the four-momentum $p^{(a)}$ of point emitter  can be obtained by the four-momentum $p^\mu$ in a rotating black hole spacetime surrounded by a CDM halo, i.e.,
\begin{eqnarray}\label{trans02}
p^{(a)}=\Lambda^{(a)}_{\;\;\;(b)}\eta^{(b)(c)}e^\mu_{\;\;(c)}p_\mu.
\end{eqnarray}
where $\eta^{(b)(c)}$ is the flat Minkowski metric. The zero-angular-momentum-observer (ZAMO) tetrad $e^\mu_{\;\;(c)}$ and the Lorentz transformation $\Lambda^{(a)}_{\;\;\;(b)}$ are respectively given by
\begin{equation}\label{tetrad}
e^\mu_{\;\;(c)}=\left[
\begin{array}{cccc}
\frac{1}{r_s}\sqrt{\frac{\Xi_s}{\Delta_s}} & 0 & \frac{\omega_s}{r_s}\sqrt{\frac{\Xi_s}{\Delta_s}} & 0 \\
0 & \frac{\sqrt{\Delta_s}}{r_s} & 0 & 0 \\
0 & 0 & \frac{r_s}{\sqrt{\Xi_s}} & 0 \\
0 & 0 & 0 & -\frac{1}{r_s}
\end{array}
\right],
\end{equation}
and
\begin{equation}\label{boost02}
\Lambda^{(a)}_{\;\;\;(b)}=\left[
\begin{array}{cccc}
\gamma & -\beta_\nu\gamma\cos\chi & -\beta_\nu\gamma\sin\chi & 0 \\
-\beta_\nu\gamma\cos\chi & (\gamma-1)\cos\chi^2+1 & (\gamma-1)\sin\chi\cos\chi & 0 \\
-\beta_\nu\gamma\sin\chi & (\gamma-1)\sin\chi\cos\chi & (\gamma-1)\sin\chi^2+1 & 0 \\
0 & 0 & 0 & 1
\end{array}
\right],
\end{equation}
where $\gamma=\frac{1}{\sqrt{1-\beta^2}}$ is the Lorentz factor. In the local frame, the temporal components of polarization vector $f^{(t)}=0$. Since the three-dimensional electric vector $\vec{E}$ of photon is along the direction of vector $\vec{p}\times\vec{B}$, the  spatial components of polarization vector $\vec{f}$ can be expressed as
\begin{eqnarray}\label{polarized01}
f^{(r)}=\frac{p^{(\phi)}\times{B^{(\theta)}}}{|\vec{p}|}, \quad\quad\quad
f^{(\phi)}=\frac{p^{(\theta)}\times{B^{(r)}}}{|\vec{p}|},\quad\quad\quad
f^{(\theta)}=\frac{p^{(r)}\times{B^{(\phi)}}}{|\vec{p}|}.
\end{eqnarray}
Thus, the photon polarization vector in the black hole spacetime can be obtained from the local polarization vector by using an inverse transformation as
\begin{eqnarray}\label{trans03}
f^\mu=e_{\;\;(c)}^\mu\Lambda^{\;\;\;(c)}_{(a)}f^{(a)},
\end{eqnarray}
where $\Lambda^{\;\;\;(c)}_{(a)}$ is the inverse matrix of $\Lambda^{(a)}_{\;\;\;(b)}$. And the photon polarization vector satisfies the relationship as
\begin{eqnarray}\label{polarized02}
f^\mu{f_\mu}=\sin\zeta^2|\vec{B}|^2.
\end{eqnarray}
Here, the $\zeta$ is the angle between photon momentum $\vec{p}$ and the magnetic field $\vec{B}$ and has the form
\begin{eqnarray}\label{angle}
\sin\zeta=\frac{|\vec{p}\times\vec{B}|}{|\vec{p}||\vec{B}|}.
\end{eqnarray}
\begin{figure}[ht]
\includegraphics[width=14cm]{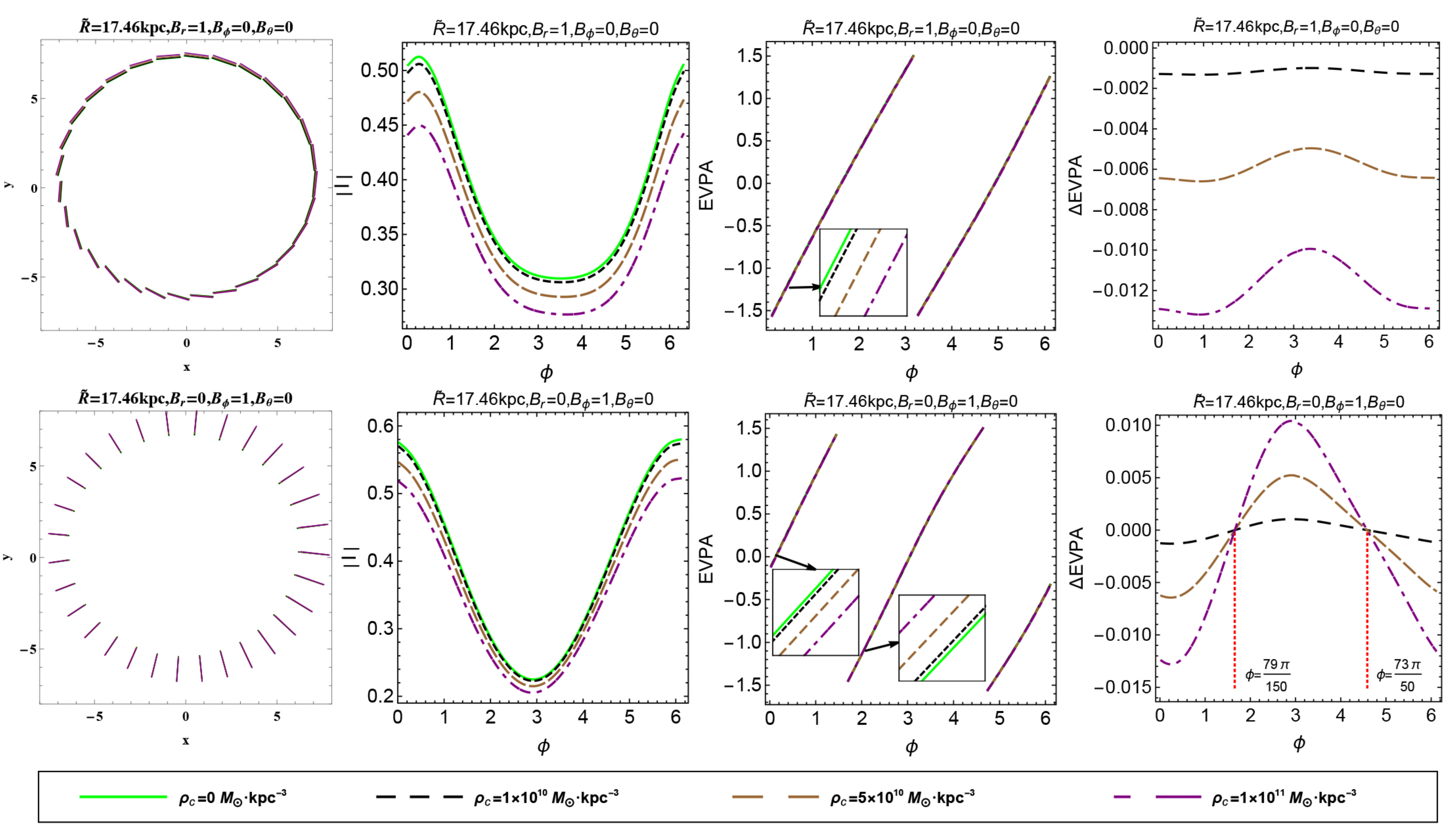}
\caption{Effects of $\rho_c$ on the polarized intensity and EVPA in the rotating black hole spacetime \eqref{metric1}. $\tilde{R}$ is fixed to be 17.46kpc. Here $r_s=6$, $a=0.3$, $\theta_o=20^{\circ}$, $\beta_\nu=0.3$, and $\chi=-90^{\circ}$.}
\label{f1}
\end{figure}
\begin{figure}[htb!]
\includegraphics[width=14cm]{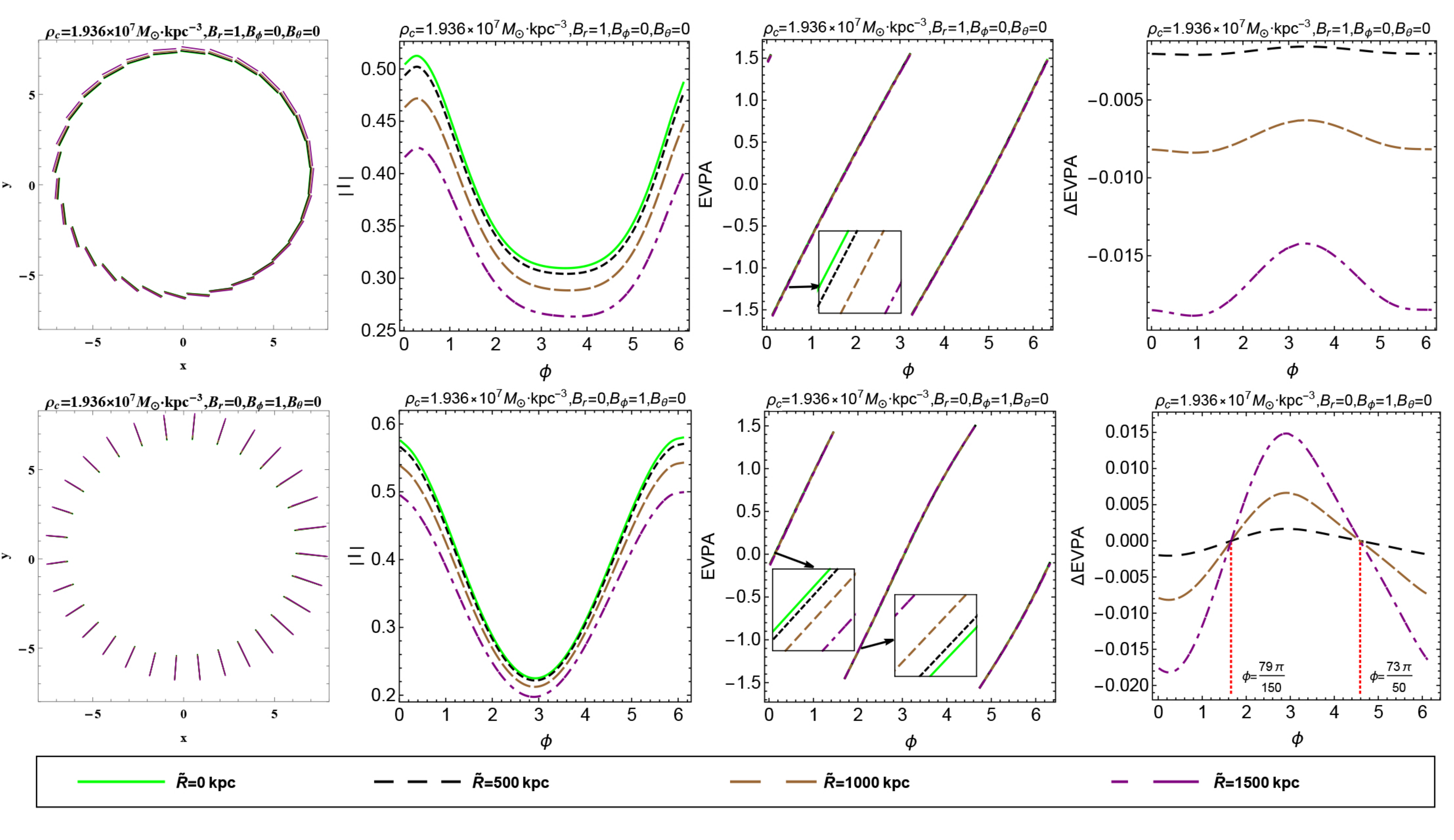}
\caption{Effects of $\tilde{R}$ on the polarized intensity and EVPA in the rotating black hole spacetime \eqref{metric1}. $\rho_c$ is fixed to be 1.936$\times10^7$ M$_{\odot}$$\cdot$kpc$^{-3}$. Here $r_s=6$, $a=0.3$, $\theta_o=20^{\circ}$, $\beta_\nu=0.3$, and $\chi=-90^{\circ}$.}
\label{f2}
\end{figure}
\begin{figure}[htb!]
\includegraphics[width=14cm]{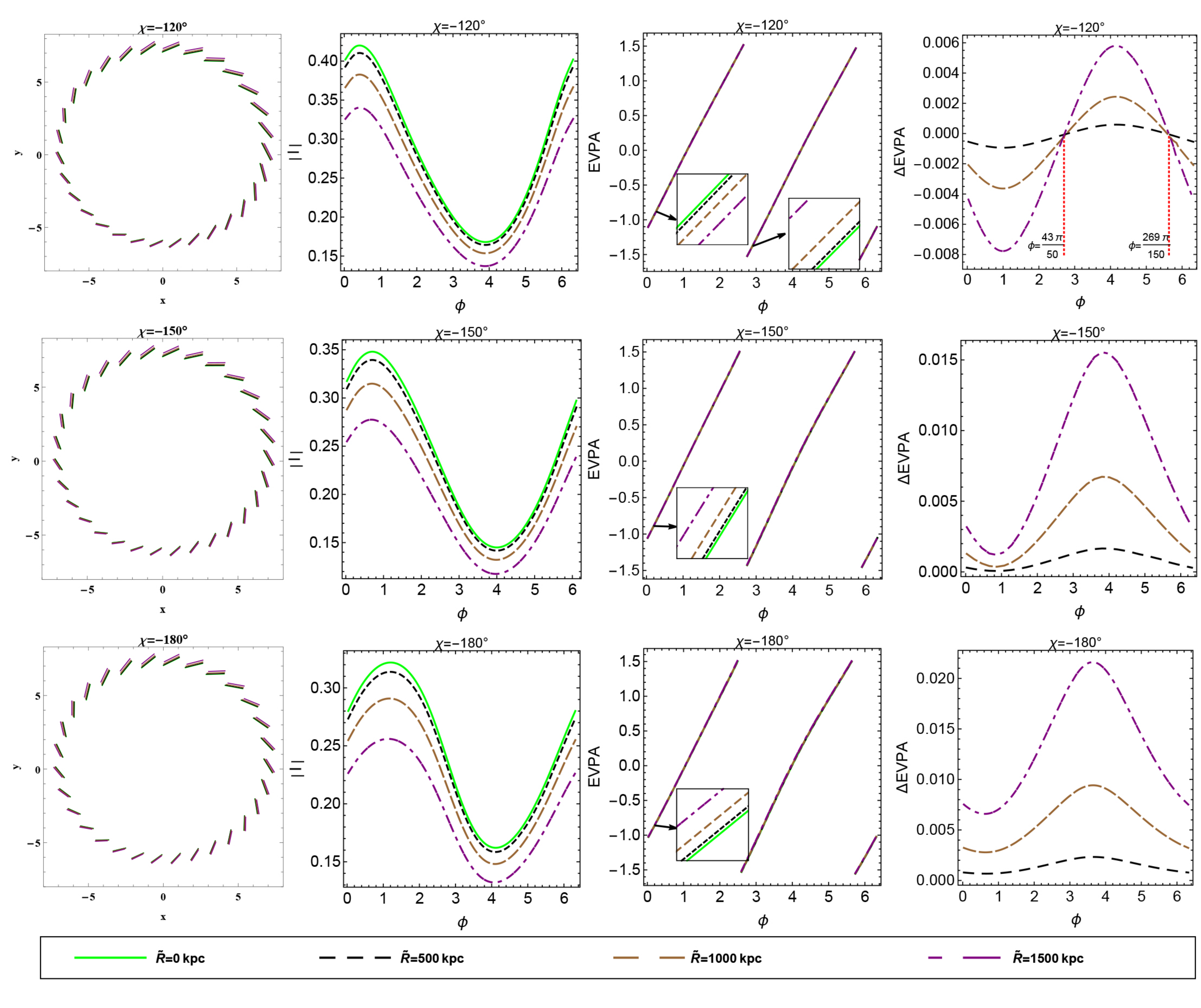}
\caption{Effects of $\tilde{R}$ on the polarized intensity and EVPA in the rotating black hole spacetime \eqref{metric1} for different fluid direction angle $\chi$. $\rho_c$ is fixed to be 1.936$\times10^7$ M$_{\odot}$$\cdot$kpc$^{-3}$. Here $r_s=6$, $a=0.3$, $\theta_o=20^{\circ}$, $\beta_\nu=0.3$, $B_r=0.87$, $B_\phi=0.5$ and $B_\theta=0$.}
\label{f3}
\end{figure}
\begin{figure}[ht]
\includegraphics[width=14cm]{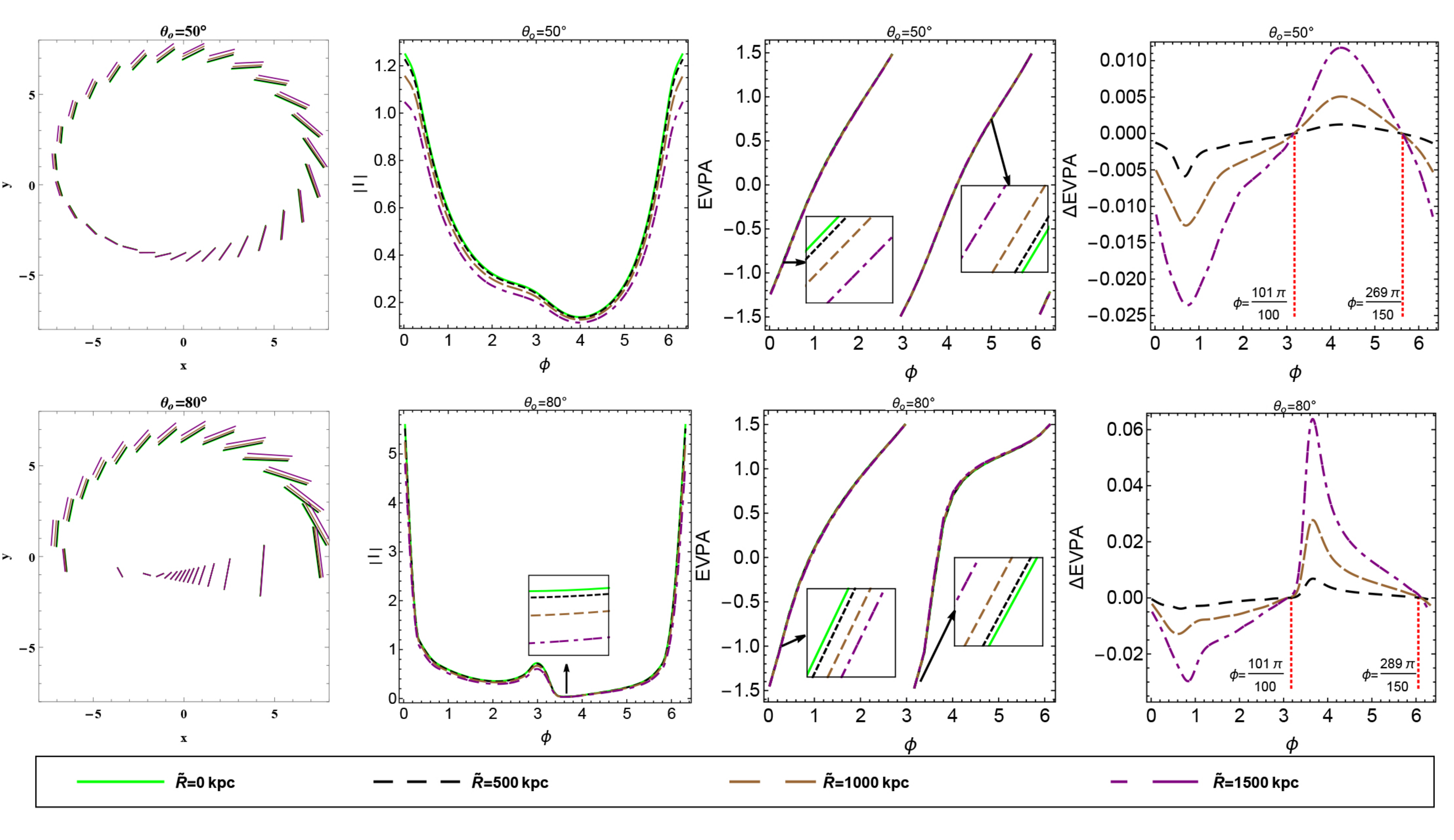}
\caption{Effects of $\tilde{R}$ on the polarized intensity and EVPA in the rotating black hole spacetime \eqref{metric1} for different observer inclination angle $\theta_o$. $\rho_c$ is fixed to be 1.936$\times10^7$ M$_{\odot}$$\cdot$kpc$^{-3}$. Here $r_s=6$, $a=0.3$, $\beta_\nu=0.3$, $\chi=-90^{\circ}$, $B_r=0.87$, $B_\phi=0.5$ and $B_\theta=0$.}
\label{f4}
\end{figure}
\begin{figure}[ht]
\includegraphics[width=14cm]{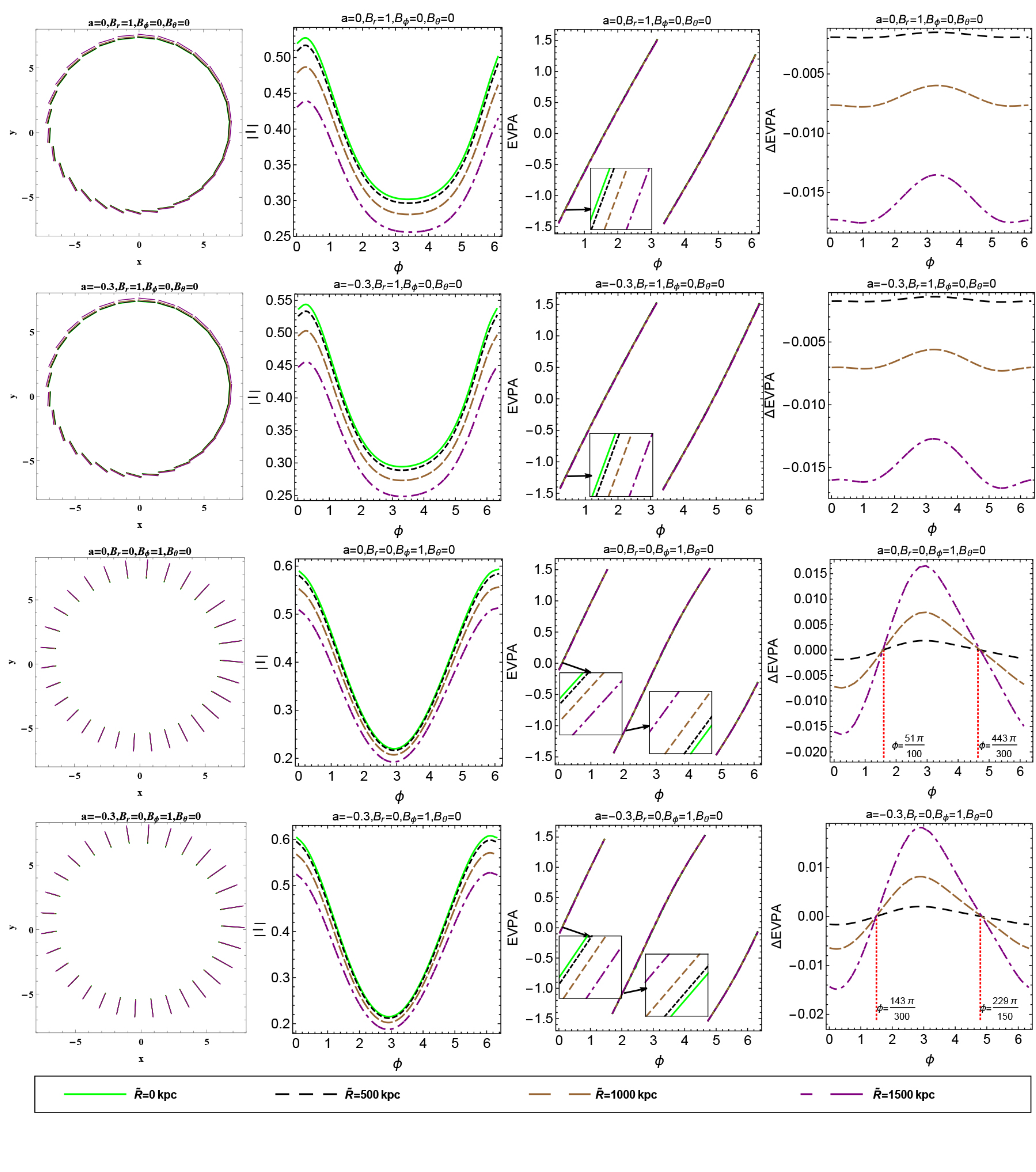}
\caption{Effects of $\tilde{R}$ on the polarized intensity and EVPA in the rotating black hole spacetime \eqref{metric1} for different spin parameter $a$. $\rho_c$ is fixed to be 1.936$\times10^7$ M$_{\odot}$$\cdot$kpc$^{-3}$. Here $r_s=6$, $\chi=-90^{\circ}$, $\theta_o=20^{\circ}$, $\beta_\nu=0.3$, $B_r=0.87$, $B_\phi=0.5$ and $B_\theta=0$.}
\label{f5}
\end{figure}
As the photon propagate along the null geodesic in the rotating black hole spacetime surrounded by a CDM halo \eqref{metric1}, the polarization vector $f^\mu$ obeys
\begin{eqnarray}\label{polarized03}
f^\mu{p_\mu}=0, \quad\quad\quad  p^\mu\nabla_\mu{f^\nu}=0.
\end{eqnarray}
Since the black hole spacetime \eqref{metric1} belong to a type-D spacetime, the conserved Penrose-Walker constant $\kappa$ can be expressed as
\begin{eqnarray}\label{PW constant01}
\kappa=p^if^j(l_in_j-l_jn_i-m_i\bar{m}_j+\bar{m}_im_j)\Psi_2^{\left(-\frac{1}{3}\right)},
\end{eqnarray}
with
\begin{eqnarray}\label{PW constant02}
&&\kappa=\kappa_1+i \kappa_2=(A-i{B})\Psi_2^{-\frac{1}{3}}, \\ \nonumber
&&A=(p^tf^r-p^rf^t)+a\sin\theta^2(p^rf^\phi-p^\phi{f^r}), \\ \nonumber
&&B=\left[(r^2+a^2)(p^\phi{f^\theta}-p^\theta{f^\phi})-a(p^tf^\theta-p^\theta{f^t})\right]\sin\theta.
\end{eqnarray}
Here $\Psi_2$ is the Weyl scalar  with the form
\begin{eqnarray}\label{weyl}
\Psi_2&=&\frac{1}{r^2+a^2\cos\theta^2}\bigg[\frac{r^2}{12}h''(r)-\frac{\left(r^2+2iar\cos\theta\right)}{6\left(r-i a \cos\theta\right)}h'(r)+\frac{\left(r^2+4iar\cos\theta-a^2\cos\theta^2\right)}{6(r-i a \cos\theta)^2}h(r)\nonumber\\
&-&\frac{6Mr+r^2+a\cos\theta(6iM+4ir-a\cos\theta)}{6(r-ia\cos\theta)^2}\bigg],
\end{eqnarray}
where $h(r)=\left(1+\frac{r}{\tilde{R}}\right)^{-\frac{8\pi{G}\rho_c\tilde{R}^3}{c^2r}}$.
The Walker-Penrose constant builds a bridge between  polarization vectors at the source and at the observer. Using the celestial coordinates $(x,y)$ and the Penrose-Walker constant $\kappa$ at the source, the polarization vector on the observer's screen can be expressed as \cite{PWconstant,Chandrasekhar,Himwich}
\begin{eqnarray}\label{polarized04}
f^x&=&\left[-1-\frac{5\xi}{12}+\frac{\xi}{2}\ln\left(\frac{R_0}{\tilde{R}}\right)\right]^{\frac{1}{3}}\frac{y\kappa_2-\mu\kappa_1}{\mu^2+y^2}, \\ \nonumber
f^y&=&\left[-1-\frac{5\xi}{12}+\frac{\xi}{2}\ln\left(\frac{R_0}{\tilde{R}}\right)\right]^{\frac{1}{3}}\frac{y\kappa_1+\mu\kappa_2}{\mu^2+y^2}, \\ \nonumber
\mu&=&-(x+a\sin\theta_o).
\end{eqnarray}
Where the quantity $\xi = -\frac{8\pi{G}\rho_c\tilde{R}^3}{c^2}$. $R_0$ is the distance between the centre of the black hole and the observer. The intensity of the linearly polarized synchrotron radiation from hot gas near the black hole to the observer can be expressed as \cite{PZ1,PZ2}
\begin{eqnarray}\label{intens01}
|I|=g^{3+\alpha_\nu}l_p|\vec{B}|^{1+\alpha_\nu}\sin\zeta^{1+\alpha_\nu},
\end{eqnarray}
Here, $g$ is the redshift factor of the photon travelling from the source to the observer. The quantity $l_p$ is the geodesic path length which the photon through the emitting material and has a form $l_p=\frac{p_s^{(t)}}{p_s^{(z)}}H$. The height of the disk $H$ can be taken be a constant $1$. The power $\alpha_\nu$ of the magnetic field $\vec{B}$ is related to the properties of the accretion disk and can be set to $\alpha_\nu=1$ \cite{PZ1,PZ2}. Then, the observed components of the polarization vector can be expressed as
\begin{eqnarray}\label{intens02}
f_{obs}^x=\sqrt{l_p}g^2|B|\sin\zeta f^x,  \quad\quad\quad f_{obs}^y=\sqrt{l_p}g^2|B|\sin\zeta f^y.
\end{eqnarray}
The total polarization intensity and the EVPA in the observer's screen can be expressed as
\begin{eqnarray}\label{intens03}
I=\left(f_{obs}^x\right)^2+\left(f_{obs}^y\right)^2,  \quad\quad\quad {\rm{EVPA}}=\frac{1}{2}\arctan\frac{U}{Q},
\end{eqnarray}
where $Q$ and $U$ are the Stokes parameters
\begin{eqnarray}\label{intens04}
Q=\left(f_{obs}^y\right)^2-\left(f_{obs}^x\right)^2,  \quad\quad\quad U=-2f_{obs}^xf_{obs}^y.
\end{eqnarray}
\begin{figure}[htb!]
\includegraphics[width=14cm]{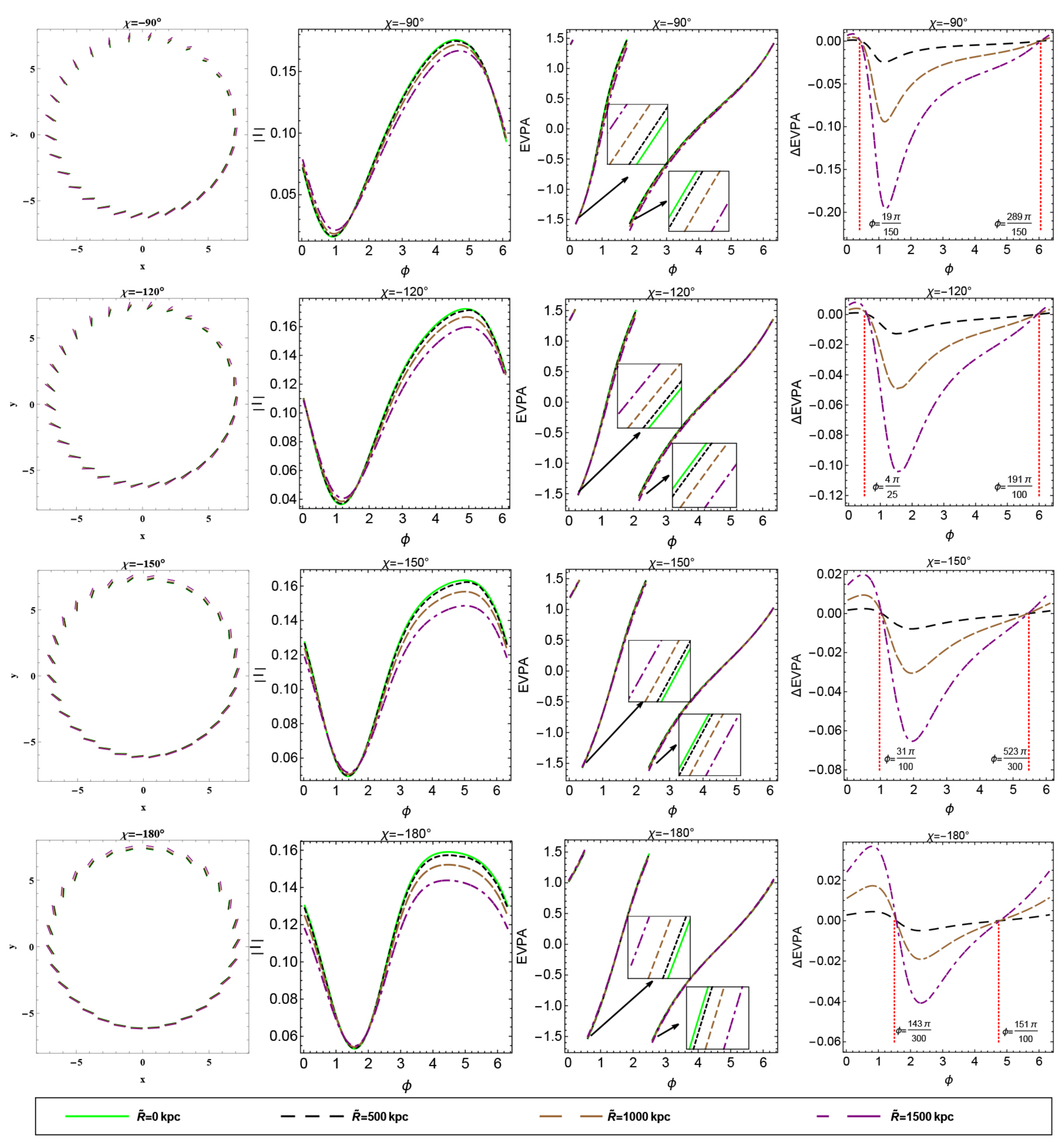}
\caption{Effects of $\tilde{R}$ on the polarized intensity and EVPA in the rotating black hole spacetime \eqref{metric1} for different $\chi$ in the case with the magnetic field owned only the vertical component $B_\theta$. $\rho_c$ is fixed to be 1.936$\times10^7$ M$_{\odot}$$\cdot$kpc$^{-3}$. Here $r_s=6$, $a=0.3$, $\theta_o=20^{\circ}$, $\beta_\nu=0.3$, $B_r=0$, $B_\phi=0$ and $B_\theta=1$.}
\label{f6}
\end{figure}
For a rotating black hole surrounded by a CDM halo \eqref{metric1}, the polarization intensity and EVPA in the pixel related to the point source can be obtained by making use of the set of celestial coordinates $(x,y)$ and Eqs. \eqref{PW constant01}, \eqref{polarized04}, \eqref{intens02}, \eqref{intens03} and \eqref{intens04}. The effects of dark matter halo on the total polarization image can be presented through repeating similar operations along the emitting ring.

\section{Effects of the CDM halo on the polarized image of a black hole}
\begin{figure}[htb!]
\includegraphics[width=14cm]{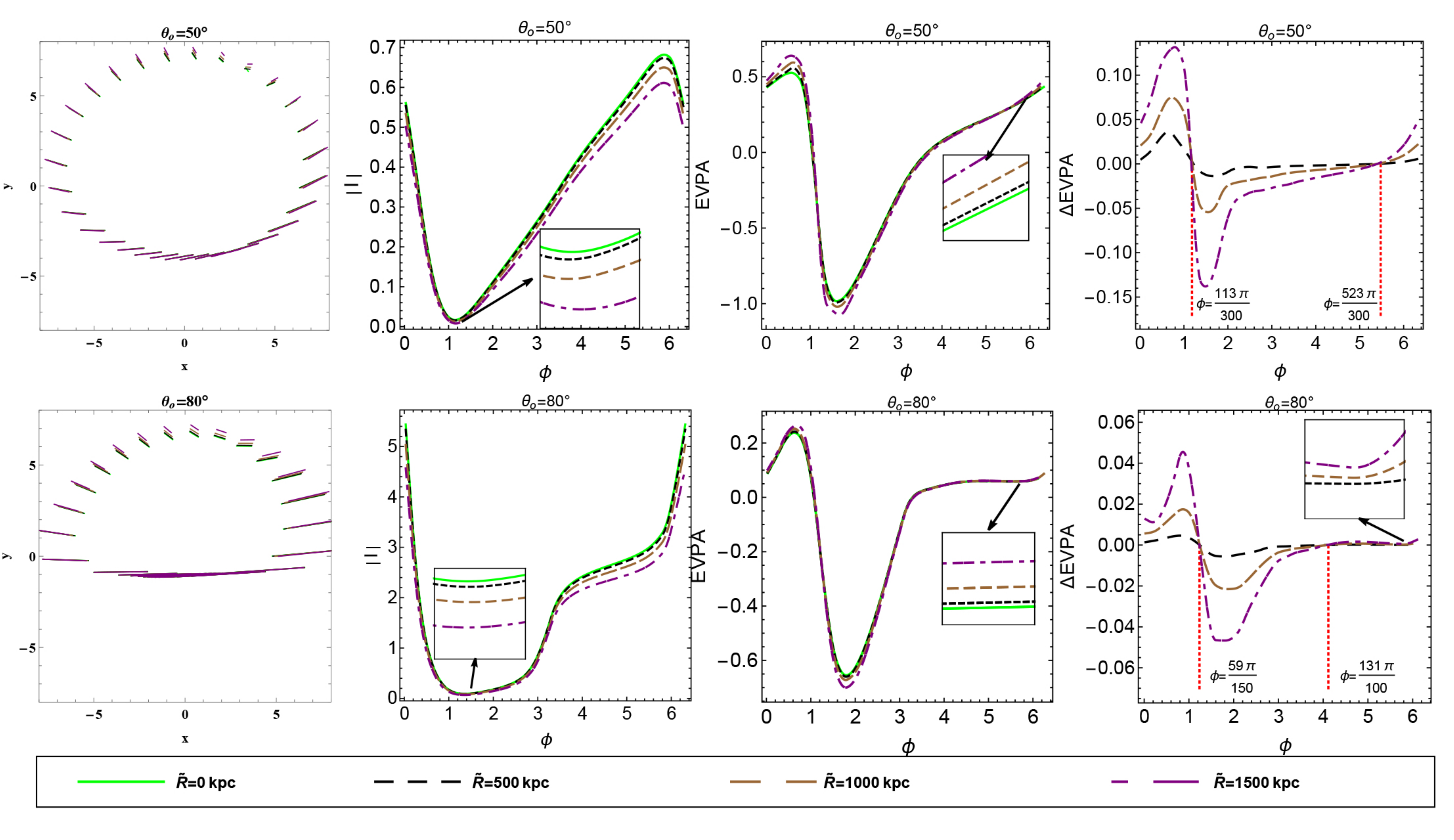}
\caption{Effects of $\tilde{R}$ on the polarized intensity and EVPA in the rotating black hole spacetime \eqref{metric1} for different $\theta_o$ in the case with the magnetic field owned only the vertical component $B_\theta$. $\rho_c$ is fixed to be 1.936$\times10^7$ M$_{\odot}$$\cdot$kpc$^{-3}$. Here $r_s=6$, $a=0.3$, $\chi=-90^{\circ}$, $\beta_\nu=0.3$, $B_r=0$, $B_\phi=0$ and $B_\theta=1$.}
\label{f7}
\end{figure}
\begin{figure}[htb!]
\includegraphics[width=14cm]{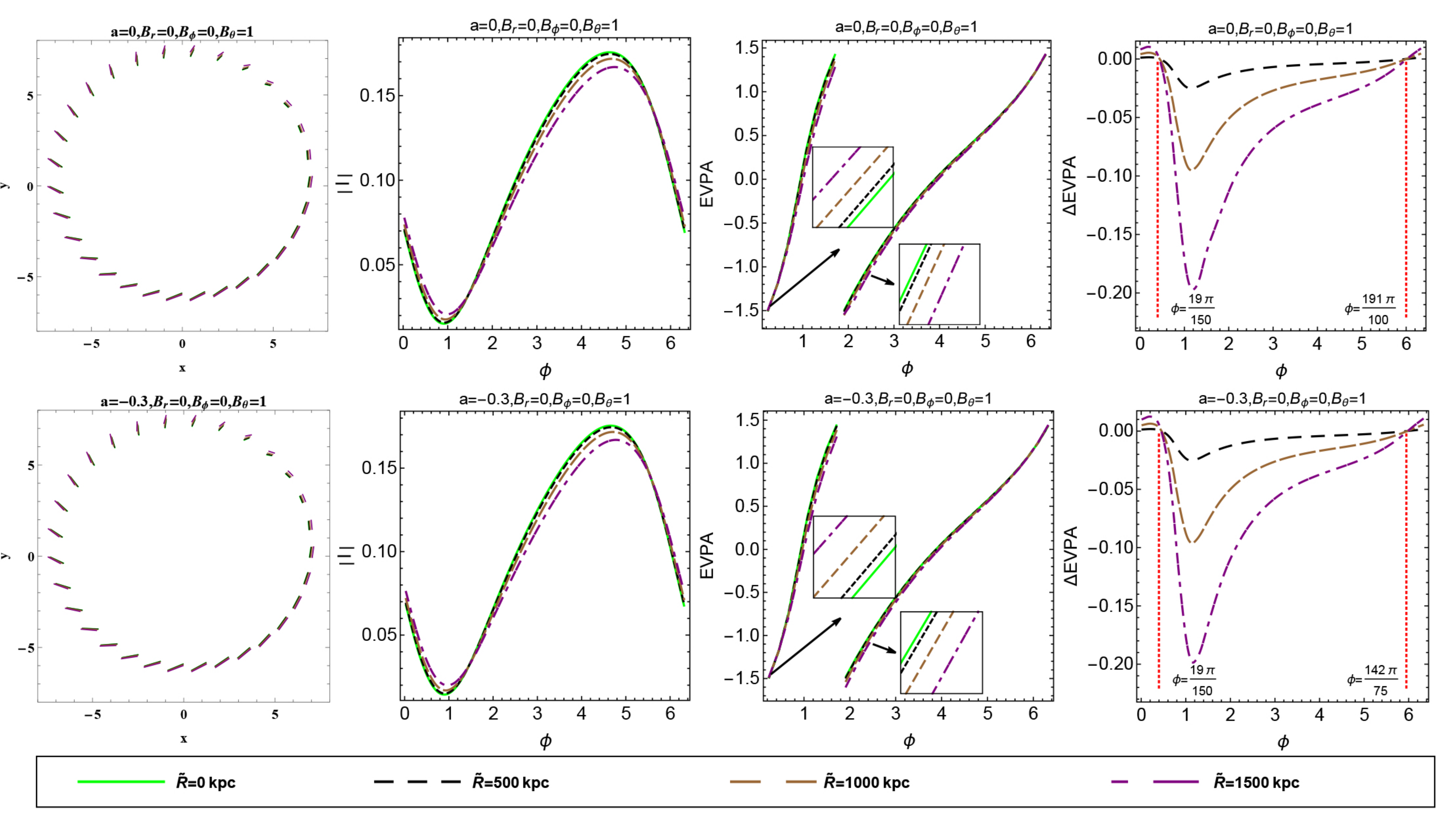}
\caption{Effects of $\tilde{R}$ on the polarized intensity and EVPA in the rotating black hole spacetime \eqref{metric1} for different $a$ in the case with the magnetic field owned only the vertical component $B_\theta$. $\rho_c$ is fixed to be 1.936$\times10^7$ M$_{\odot}$$\cdot$kpc$^{-3}$. Here $r_s=6$, $\chi=-90^{\circ}$, $\theta_o=20^{\circ}$, $\beta_\nu=0.3$, $B_r=0$, $B_\phi=0$ and $B_\theta=1$.}
\label{f8}
\end{figure}
Now, we can probe the effects of the CDM halo on the polarization image of the equatorial emitting ring with radius $r_s=6$ around the rotating black hole \eqref{metric1}.
Figs.(\ref{f1}) and (\ref{f2}) show the effects of the density $\rho_c$ and the characteristic radius $\tilde{R}$ of CDM halo on the polarized image of the emitting ring in the case where  the magnetic field lies in the equatorial plane. With increasing $\rho_c$ or $\tilde{R}$,  the polarization intensity decreases, while the EVPA decreases in the case with pure radial magnetic field and the change of the EVPA depends on the coordinate $\phi$ in the case of the pure angular magnetic field. In other words, the effects of the density $\rho_c$ are similar to those of the characteristic radius $\tilde{R}$. It can be explained by a fact that the metric function $f(r)$ in Eq. \eqref{metric1} decreases with $\rho_c$ and $\tilde{R}$.
Thus, in the following parts, we only discuss the effects of the characteristic radius $\tilde{R}$ of the CDM halo on the polarized image.

Figs.(\ref{f3})-(\ref{f5}) present respectively the effects of the radius $\tilde{R}$ of the CDM halo on the polarized image in the case the magnetic field lies in the equatorial plane for different fluid direction angles $\chi$,  observer inclination angles $\theta_o$ and spin parameters of the black hole.
When $\chi$ changes from $-120^{\circ}$ to $-180^{\circ}$, the polarization intensity still decreases with $\tilde{R}$,  the change of the EVPA becomes more complicated. The region where the EVPA increases with $\tilde{R}$ becomes gradually broader with $\chi$, so that the EVPA finally becomes a increasing function of $\tilde{R}$. This is also shown in the change of the quantity $\Delta{\rm EVPA}\equiv{\rm EVPA-EVPA_{K}}$, where $\rm EVPA_{K}$ denotes the corresponding EVPA in the Kerr black hole spacetime. Similarly, with the increase of the observer inclination angle $\theta_o$, the polarization intensity decreases with $\tilde{R}$ and the region where the EVPA increases with $\tilde{R}$ becomes gradually broader. However, the EVPA finally does not become a increasing function of $\tilde{R}$ in the high spin case. Moreover, one can find that in the rotating black hole case the dependence of the polarization intensity and EVPA on the CDM halo radius $\tilde{R}$ is
qualitatively similar to that in the non-rotating case.

Figs.(\ref{f6})-(\ref{f8}) also present the effects of $\tilde{R}$ on the polarized image in the case where the magnetic field is perpendicular to the equatorial plane. For different $\chi$, the changes of the polarization intensity are different from those in the case with the pure equatorial magnetic field.
As the $\chi$ changes from $-90^{\circ}$ to $-180^{\circ}$, the region where the polarized intensity decreases with $\tilde{R}$ becomes broad, while the region where EVPA decrease with $\tilde{R}$ becomes narrow.
However, in the case of $\chi=-90^{\circ}$, the decreasing of EVPA with $\tilde{R}$ is dominated.
With the increase of $\theta_o$, the dependence of the  polarization intensity on the angular coordinate $\phi$ is changed, but the polarization intensity still decreases with $\tilde{R}$. The region where EVPA decrease with $\tilde{R}$ becomes narrow, which is similar to that in the case where the $\chi$ changes from $-90^{\circ}$ to $-180^{\circ}$.
Fig.(\ref{f8}) shows the changes of the polarization intensity and EVPA with $\tilde{R}$ are also qualitatively similar for the different spin parameter $a$ in this case.
\begin{figure}[htb!]
\includegraphics[width=13cm]{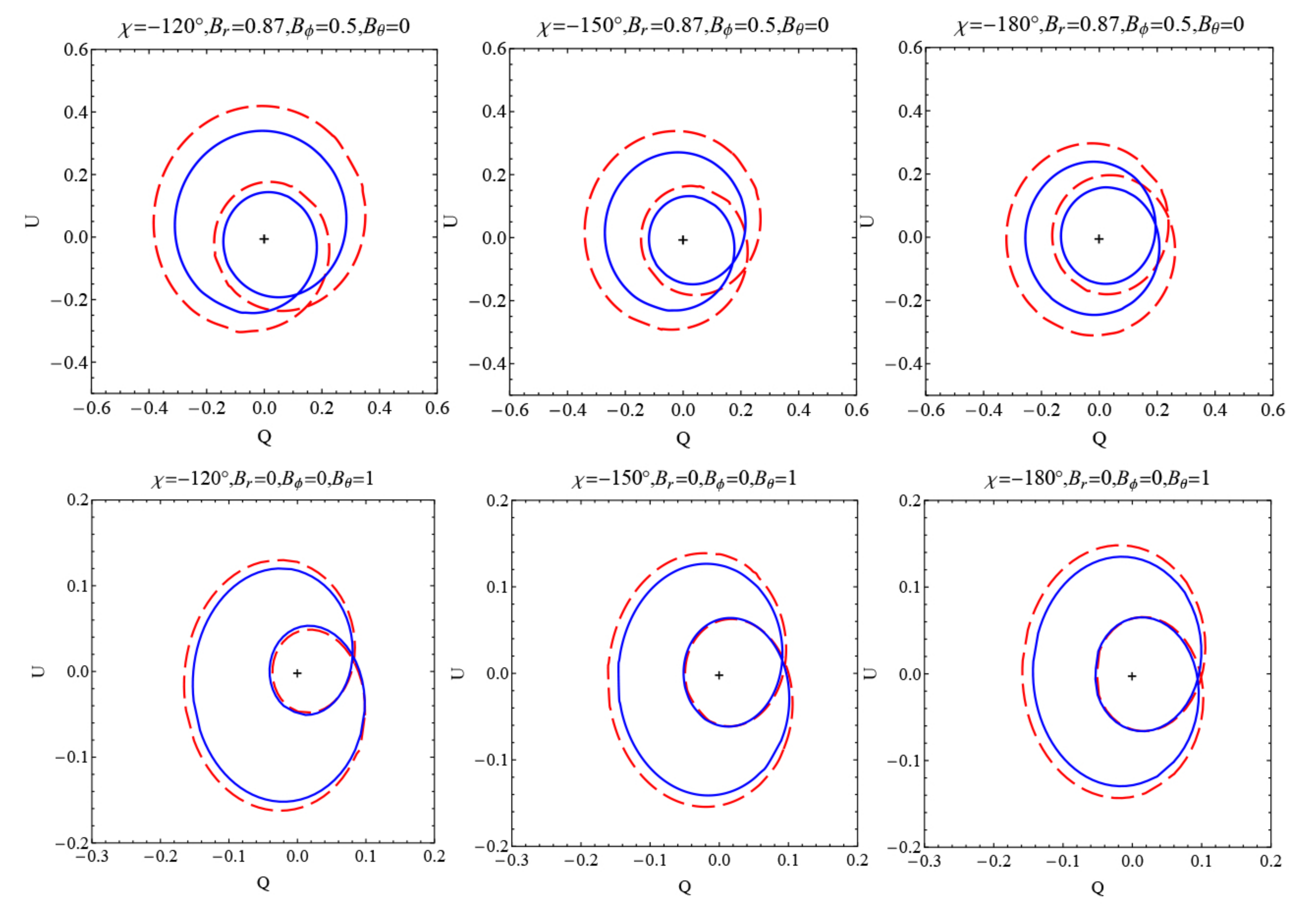}
\caption{Effects of $\tilde{R}$ on the $Q-U$ diagram for different $\chi$ in the rotating black hole spacetime \eqref{metric1}. Here $r_s=6$, $\theta_o=20^{\circ}$, $a=0.3$, and $\beta_\nu=0.3$. As $\rho_c$ is fixed to be 1.936$\times10^7$ M$_{\odot}$$\cdot$kpc$^{-3}$. The red solid line and the blue dashed line correspond to the cases with the $\tilde{R}=0$ kpc and $\tilde{R}=1500$ kpc, respectively. Black crosshairs indicate the origin of each plot.}
\label{f9}
\end{figure}
\begin{figure}[htb!]
\includegraphics[width=13cm]{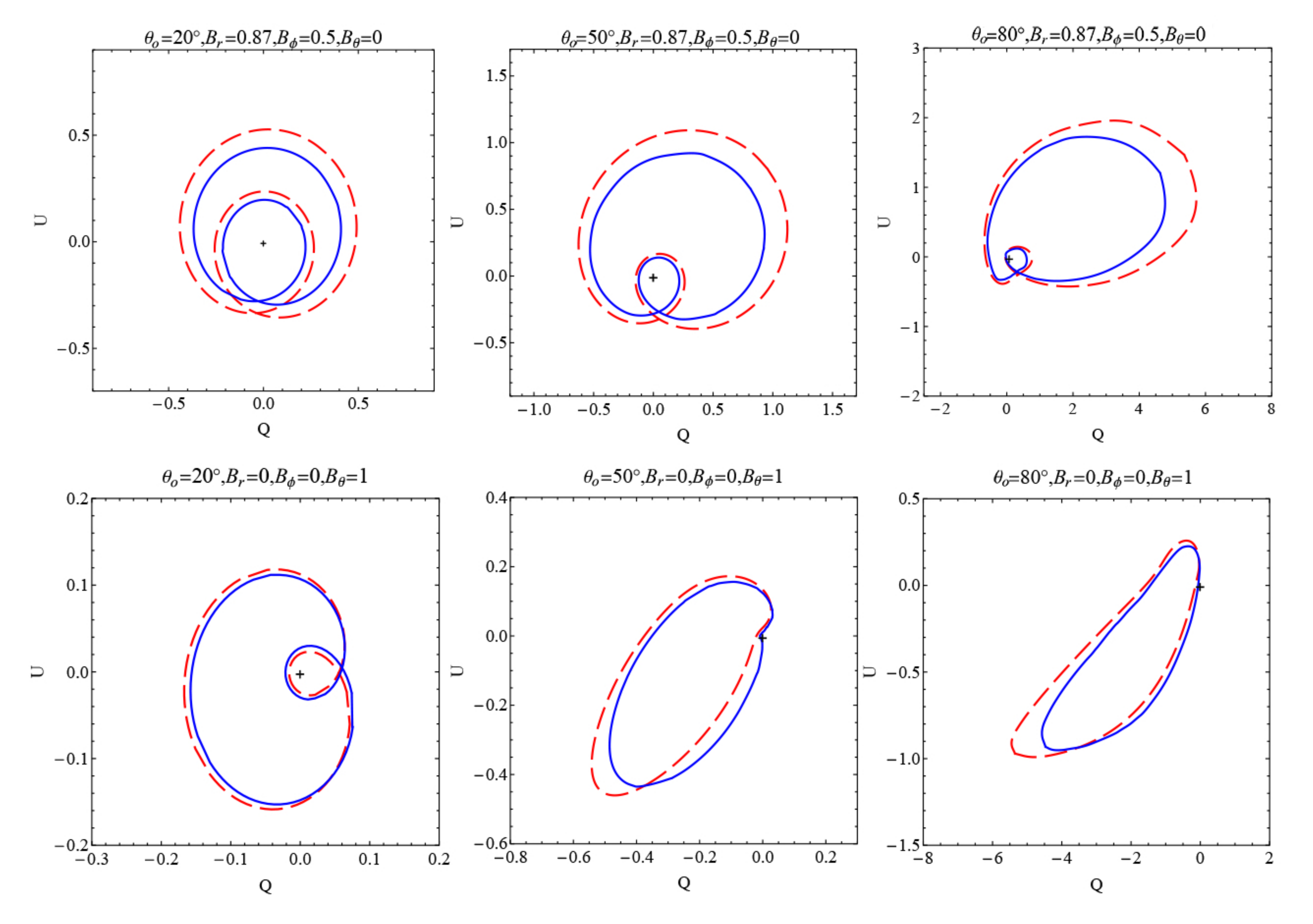}
\caption{Effects of $\tilde{R}$ on the $Q-U$ diagram for different $\theta_o$ in the rotating black hole spacetime \eqref{metric1}. Here $r_s=6$, $\chi=-90^{\circ}$, $a=0.3$, and $\beta_\nu=0.3$. As $\rho_c$ is fixed to be 1.936$\times10^7$ M$_{\odot}$$\cdot$kpc$^{-3}$. The red solid line and the blue dashed line correspond to the cases with the $\tilde{R}=0$ kpc and $\tilde{R}=1500$ kpc, respectively. Black crosshairs indicate the origin of each plot.}
\label{f10}
\end{figure}

Finally, Figs.(\ref{f9})-(\ref{f10}) show the effects of $\tilde{R}$ on the $Q-U$ loops patterns in the image of the emitting ring around a rotating black hole surrounded by the CDM halo, which depend heavily on the magnetic field configuration, the fluid velocity, the observation inclination angle and the spin parameter of the black hole.
In the case with the lower observed inclination, the size of the two loops decreases with the increase of $\tilde{R}$  as the magnetic field lies in
the equatorial plane, while in the case of the magnetic field being perpendicular to the equatorial plane, the size of the outer loop decreases and the inner loop increases. For the case of the magnetic field lies in the equatorial plane,
the sizes of two loops decrease with $\tilde{R}$ although the inner loop dramatically shrinks in the higher observed inclination case. For the case where the direction of magnetic field is perpendicular to the equatorial plane, the size of the outer loops decreases and the inner loops increases with $\tilde{R}$  in the lower observed inclination $\theta_o=20^{\circ}$. However, in the higher inclination angle case, the inner loop vanishes and the change of loop size with the
CDM halo parameter $\tilde{R}$ becomes more complicated.

\section{Summary}

We have studied the polarized image of an equatorial emitting ring around the rotating black hole surrounded by a CDM halo. Results show that the effects of the CDM halo density $\rho_c$  on the polarization intensity and EVPA are similar to those of the characteristic radius $\tilde{R}$ of the halo. The changes of the polarization intensity and EVPA with $\tilde{R}$ also depend on the magnetic field configuration, the fluid velocity and the observed inclination angle in the spacetime  of the rotating black hole surrounded by a CDM halo. When the magnetic field lies in the equatorial plane, with the increase of $\tilde{R}$, the polarization intensity decreases, but in the case of the pure radial magnetic field, the EVPA only decreases  and the change of the EVPA depends on the coordinate $\phi$. When the angle $\chi$ changes from $-90^{\circ}$ to $-180^{\circ}$, the region where the EVPA increases with $\tilde{R}$ becomes gradually broader, so that the EVPA finally becomes a increasing function of $\tilde{R}$. When the magnetic field is perpendicular to the equatorial plane, as the $\chi$ changes from $-90^{\circ}$ to $-180^{\circ}$, the region where the polarized intensity decreases with $\tilde{R}$ becomes broad, while the region where EVPA decrease with $\tilde{R}$ becomes narrow.
However, in the case of $\chi=-90^{\circ}$, the decreasing of EVPA with $\tilde{R}$ is dominated.
With the increase of $\theta_o$,  the polarization intensity  also decreases with $\tilde{R}$, and the region where the EVPA decreases with $\tilde{R}$ becomes narrow.

We also present the effects of $\tilde{R}$ on the $Q-U$ loops,  which depend heavily on the magnetic field configuration, the fluid velocity, the observation inclination angle and the spin parameter of the black hole.
In the case with the lower observed inclination, the size of the two loops decreases with the increase of $\tilde{R}$  as the magnetic field lies in
the equatorial plane, and the size of the outer loop decreases and the inner loop increases as the magnetic field is vertical to the equatorial plane. In the higher inclination angle case, the inner loop vanishes and the change of loop size with the
CDM halo parameter $\tilde{R}$ becomes more complicated. Moreover, for the fixed $\tilde{R}$ or $\rho_c$, the dependence of polarization image on the magnetic field configuration, the fluid velocity and the observed inclination angle in rotating black hole surrounded by a CDM halo are similar to those in other rotating black hole cases.

Finally, our results also indicate that the influence
of the CDM halo on the polarized image of the emitting ring around the black hole \eqref{metric1} is generally minor, which is consistent with the effects of dark matter halo on  black hole shadows \cite{CDM2,CDM201}. This implies that the detection of these effects from the CDM halo is out of the reach of the current astronomical instruments. With the increasing accuracy and resolution of the future astronomical observations and the technological development, it is expected that these effects of the CDM halo on the polarize image of black holes can be detected.

\section{\bf Acknowledgments}

This work was supported by the National Natural Science
Foundation of China under Grant No.12275078 and 12035005.

\end{document}